\def\vecr{\vec r}
\def\vecp{\vec p}

\documentclass{camera}
\usepackage{graphicx}\usepackage{epsfig}\usepackage{portland}

\begin{document}

%
\title{Spinodal instability growth\\ in new stochastic approaches}

%
\author{P. Napolitani$^1$, M. Colonna$^2$ \and V. de la Mota$^3$}

%
\organization{\vspace{-1.5cm}
  $^1$ IPN, CNRS/IN2P3, Universit\'e Paris-Sud 11, 91406 Orsay cedex, France
\\$^2$ INFN-LNS, Laboratori Nazionali del Sud, 95123 Catania, Italy
\\$^3$ SUBATECH, EMN-IN2P3/CNRS-Universit\'e de Nantes, 44307 Nantes, France 
}
\maketitle
\vspace{-0.5cm}
\begin{abstract}
	Are spinodal instabilities the leading mechanism in the fragmentation
of a fermionic system?
Numerous experimental indications suggest such a scenario and
stimulated much effort in giving a suitable description, without being
finalised in a dedicated transport model.

On the one hand, the bulk character of spinodal behaviour requires an accurate
treatment of the one-body dynamics, in presence of mechanical
instabilities.
On the other hand, pure mean-field 
implementations 
do not apply to situations where
instabilities, bifurcations and chaos are present.
The evolution of instabilities should be treated in a large-amplitude
framework requiring fluctuations of Langevin type.

We present new stochastic approaches constructed by requiring a
thorough description of the mean-field response in presence of
instabilities.
Their particular relevance is an improved description of the spinodal
fragmentation mechanism at the threshold, where the instability growth
is frustrated by the mean-field resilience.
\end{abstract}
 
\section{Mechanical instabilities in one-body dynamics}

	The phase-space dynamics of fermionic systems may evolve towards inhomogeneous density patterns when mechanical instabilities set in.
	This is a general process which characterises Fermi liquids at low densities, involving in some specific conditions spinodal mechanisms of amplification of the unstable modes.
	In particular, the action of the spinodal process on isoscalar modes was suggested as a possible channel of fragment formation in heavy-ion collisions, when the system explores densities which are below nuclear saturation~\cite{Chomaz2004}.
	It may as well have some relevance in the physics of compact stars, when portions of inhomogeneous nuclear matter at sub-saturation density and at large temperatures are formed~\cite{Ducoin2007}.
%
%

	Fermionic systems in presence of instabilities, and especially in conditions which correspond to dissipative heavy-ion collisions (i.e. in proximity of Fermi energies), can be described by the competition of two processes: the growth of instabilities and the antagonist effect of the mean field.
	Therefore, in order to describe such conditions, 
the following consideration should be accounted for: while a one-body description of bulk properties is needed, pure mean field equations do not hold when instabilities set in.
	Along this direction, a transport description of the system can be constructed as based on a one-body Hamiltonian $H$ supplemented by a fluctuating contribution which adds the unknown $N$-body correlations.
	We follow the approach of taking the Wigner-transform analogue in terms of one-body distribution functions $f(\vecr,\vecp,t)$ (function of time $t$, space $\vecr$ and momentum $\vecp$ coordinates), which yields the Boltzmann-Langevin equation 
\begin{equation}
	\dot{f} = \partial_t\,f - \left\{H[f],f\right\} = {\bar{I}[f]}+{\delta I[f]} \;.
\label{eq1}
\end{equation}
The above transport equation gives the evolution of the semiclassical one-body distribution function $f$ in its own self-consistent mean field as resulting from two contributions.
	$\bar{I}[f]$ is the average hard two-body collision integral in terms of the one-body distribution function $f$, related to the mean number of transitions 
between phase-space elementary volumes 
d$\nu$.
	$\delta I[f]$ is a Markovian term, acting as a fluctuating force while conserving single-particle energies.
Also 
$\delta I[f]$ 
is expressed in terms of the one-body distribution function $f$, through its correlation, 
$\langle \delta I(\vecr,\vecp,t) \delta I(\vecr'\!,\vecp'\!,t')\rangle = 2D(\vecr,\vecp;\vecr'\!,\vecp'\!,t')\delta(t-t')$,
which contains a diffusion coefficient $D$ also related to d$\nu$ \cite{Colonna1994}.
The fluctuating term $\delta I[f]$ acts during the whole temporal evolution of the process and introduces fluctuations by exploiting $N$-body correlations.

	In order to construct a transport model, we exploit two numerical strategies to solve eq.~\ref{eq1}
which have been successfully applied to heavy-ion collisions.
	Either fluctuations are imposed from an external stochastic force (related to an external potential $U_{\textrm{ext}}$) and projected on the density space so that 
$\delta I[f]=\partial_{\vecr}\,U_{\textrm{ext}}\;\partial_{\vecp}\,f$: this is the Stochastic Mean Field~\cite{Colonna1998} strategy (SMF).
	Or, fluctuations are introduced in full phase space from inducing $N\!-\!N$ correlations: in this approach, followed in the Boltzmann-Langevin One-Body model~\cite{Napolitani2013, Napolitaniproceedings} (BLOB), each single collision event acts on a larger portion of phase space as compared to SMF (and the transition rate is scaled accordingly), with the constraint that the fluctuation amplitude is determined at equilibrium by an occupancy variance equal to $f(1-f)$ in a phase-space cell $h^3$. Correspondingly, the final states of collision events are adapted to the vacancy profile so that the Pauli principle is never violated.

\section{Mean-field response in nuclear matter}
%
%
%
\begin{figure}[b]
\includegraphics[width=0.55\textwidth]{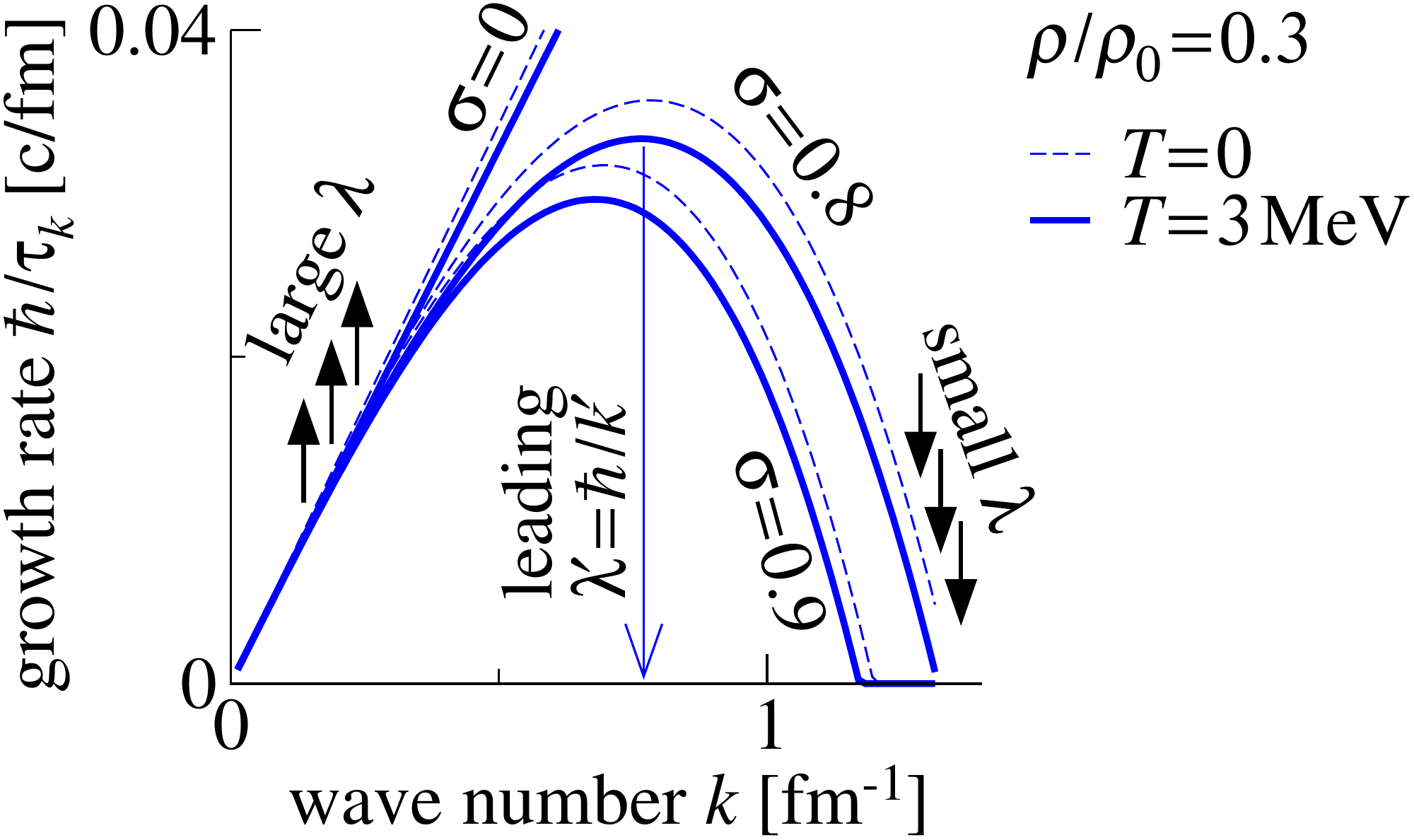}\hspace{0.05\textwidth}%
\begin{minipage}[b]{0.4\textwidth}\caption{\label{fig_dispersionrelationcartoon}Evolution of the growth rate of a disturbance as a function of the wave number in different situations. See text.}
\end{minipage}
\end{figure}

	The stochastic transport approaches based on the Boltzmann-Langevin equation which were introduced above have the aim of describing dissipative heavy-ion collisions.
	They are expected to let fluctuations develop spontaneously and 
to exhibit the correct growing in amplitude of the unstable modes as a function of time and thermodynamic conditions.	
	In order to analyse these features, these models should first of all be tested on nuclear matter in unstable conditions, so as to compare with analytical solutions of the dispersion relation for the propagation of density waves in Fermi liquids.

	Let us situate the system in a mechanically unstable region of the equation of state, choosing a temperature 
$T\!\approx 3$MeV 
and a mean-density-to-saturation-density ratio $\rho/\rho_0\!=\!0.3$ (the details of the interaction are given in ref.~\cite{Baran2005}).
	Such conditions correspond to a negative incompressibility $\chi^{-1}=\rho\,\partial_{\rho}P<0$
and to imaginary solutions of the dispersion relation, characterising a process of amplification of a disturbance in density space~\cite{Colonna1994}. 
This latter manifests as an undulation in density space of wavelength $\lambda$ and wave number $k$;
the instability characteristic time is $\tau$. 
	Fig.~\ref{fig_dispersionrelationcartoon} presents the evolution of the growth rate $\hbar/\tau$ of a disturbance as a function of the wave number $k$ for nuclear matter in different situations calculated analytically in a linear-response approximation for zero temperature and for a finite temperature $T\!=\!3$MeV.
	In particular, zero-sound conditions are studied as a function of the interaction range: this latter is introduced in the dispersion relation by applying a gaussian smearing factor of the mean-field potential (in the figure, $\sigma$ indicates the width in fm).
	For a zero-range interaction ($\sigma\!=\!0$) a linear behaviour indicates that the more matter is to be relocated the longer it takes, i.e. the larger is the characteristic time $\tau$, and it tends therefore to a growth rate equal to zero for large wavelengths (i.e. for $k\!\rightarrow\!0$).
	The introduction of a finite range of the interaction has the consequence that the growth rate drops to zero in correspondence to a largest wave number; beyond this value, the small wavelengths (i.e. large $k$) are suppressed. 
	The leading disturbance $\lambda^{\prime}$, gives the largest growth rate.
	Differently from the analytical calculation which assumes that all $k$ modes are decoupled, in a transport calculation where fluctuations appear spontaneously, small wave numbers may combine into large wave numbers: the resulting recombination of the $k$ modes modifies therefore the analytical distribution in the direction indicated by the arrows in fig.~\ref{fig_dispersionrelationcartoon}.
%
%
\begin{figure}[b!]\begin{center}
\includegraphics[angle=0, width=1\textwidth]{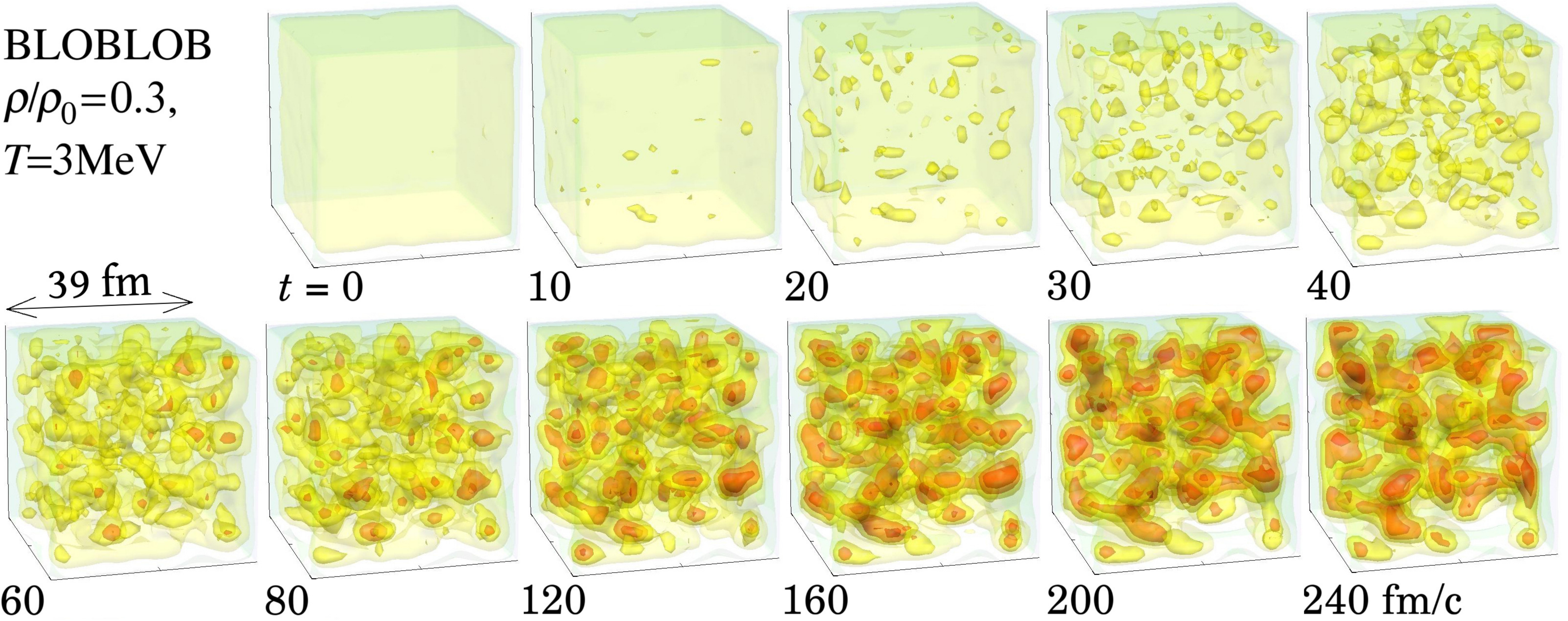}
\end{center}\caption
{
	BLOBLOB calculation. One block with boundary conditions is shown.
}
\label{fig_box}
\end{figure}

	The above description of the dispersion relation based on an analytical solution should correspond to a transport calculation applied to the same conditions.
	For this purpose, we prepared a cubic box of edge length $l\!=\!39$fm with periodic boundary conditions containing  1584 neutrons and 1584 protons (each one represented by 40 test particles), so that $\rho/\rho_0\!=\!0.3$, and at a temperature of 3~MeV.
	Fig.~\ref{fig_box} shows the evolution in time of the corresponding system as described within the BLOBLOB model, which is the BLOB approach applied to one block with periodic boundary conditions.
	In this calculation, the Boltzmann-Langevin term agitates the density profile over several $k$ waves spontaneously so that, as a function of time, a mottling pattern appears.
%
%
\begin{figure}[b!]\begin{center}
\includegraphics[angle=0, width=.9\textwidth]{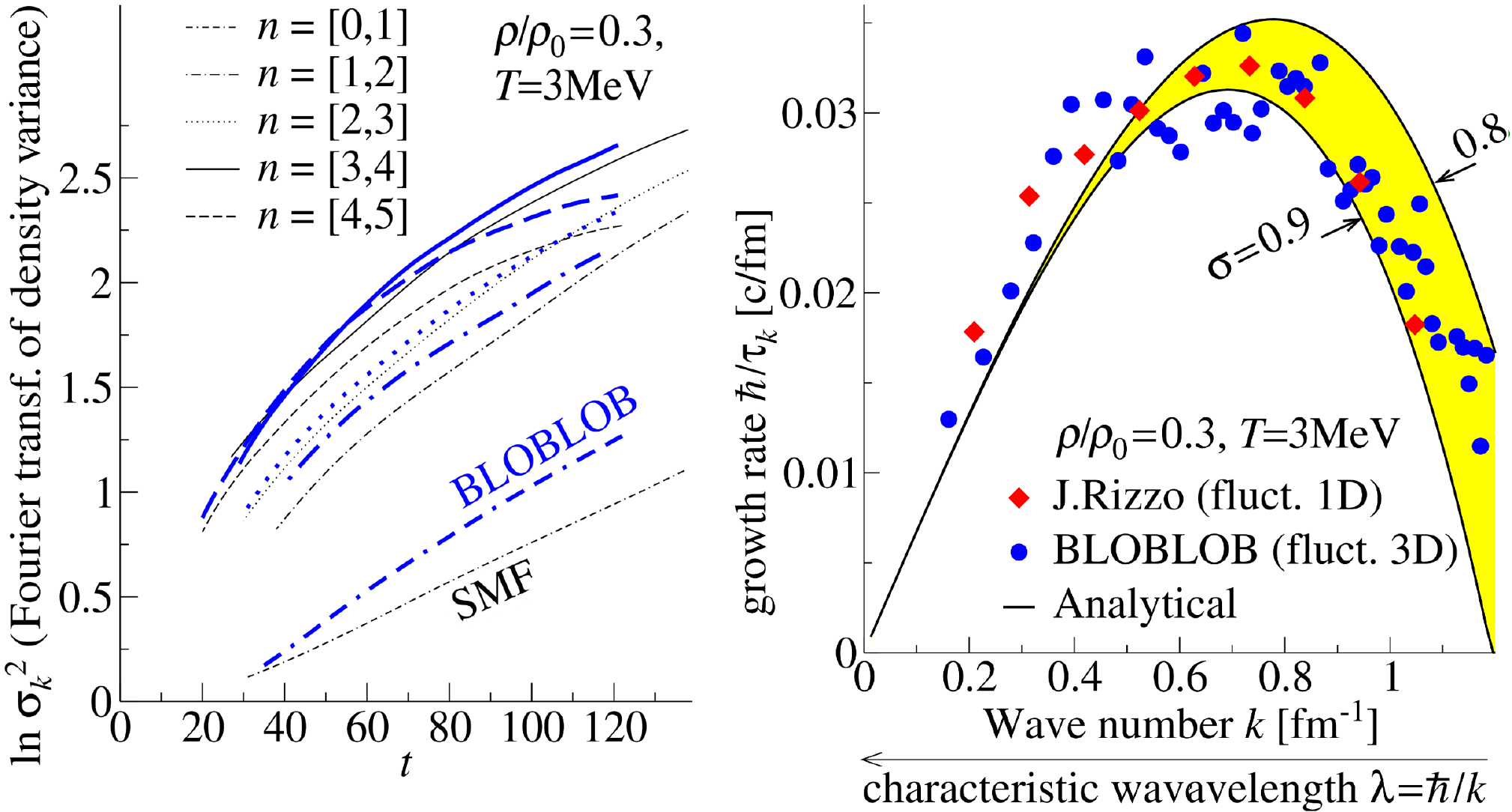}
\end{center}\caption
{
	Left. BLOBLOB (thick lines) and SMF (thin lines) calculations: Fourier transform of density 
variance 
as a function of time for disturbances of different number of nodes $n$.
	Right. Dispersion relation from the BLOBLOB calculation, compared to a calculation in one dimension and to the analytical solution.
}
\label{fig_dispersionrelation}
\end{figure}

	The growth of the amplitude of the corresponding disturbances for different wavelengths is studied as a function of time in the left panel of fig.~\ref{fig_dispersionrelation}; each curve of the figure shows in particular the Fourier transform of the density 
variance 
$\sigma^2_k$ as a function of time for a set of wavelengths which correspond to a given range of nodes of the corresponding undulation in the periodic block. The calculations are performed with the BLOBLOB model and with the SMF model also adapted to the same periodic box and with the same parameters.
	Both approaches describe the growth of unstable modes but the first one develops fluctuations more rapidly.
	The right panel of fig.~\ref{fig_dispersionrelation} shows the corresponding dispersion relation from the BLOBLOB calculation (constructed from the slope of the curves of 
ln$\sigma^2_k$ for specific $k$ values in regions showing the more linear-like behaviour).
	The dispersion relation is also calculated in a corresponding one-dimensional approach~\cite{Rizzo2008} and it is compared to the expected analytical behaviour (the band on the analytical calculation represents the interval of the gaussian smearing of the mean-field potential which is consistent with the width of the triangular functions used in the numerical calculation);
a close agreement is found between the transport calculation and the analytical expectation and the deviations on the side of the distributions are explained by the coupling of small wavelengths into large wavelengths as indicated in fig.~\ref{fig_dispersionrelationcartoon}
by the arrows. 

\section{Spinodal phenomenology in a finite system}
%
%
\begin{figure}[b!]\begin{center}
\includegraphics[angle=0, width=1\textwidth]{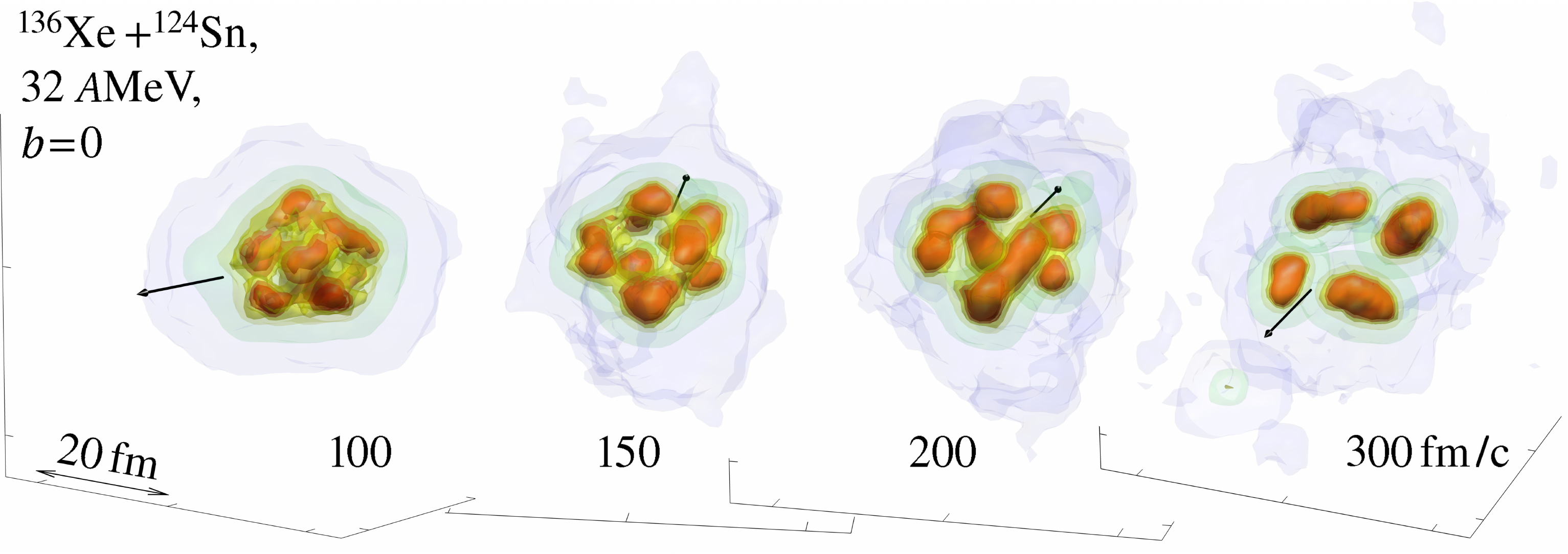}
\end{center}\caption
{
	BLOB calculation of a spinodal fragmentation process followed by a partial coalescence of the spinodal pattern into a smaller number of fragments. For better visibility, axis orientation is modified for each frame; arrows indicate the reaction axis.  
}
\label{fig_spinodal}
\end{figure}

	The calculations of the dispersion relation in nuclear matter confirm that the BLOB approach is reliable in reproducing the correct fluctuation amplitude in unstable conditions.
	We can then apply the transport model to a finite system, where the general phenomenology related to the spinodal region of the equation of state should de described~\cite{Borderie2008}.
	Outside of this region, either the system is not enough excited and diluted, so that a damped dynamics leads to the formation of a compound nucleus, or the system is very hot and very diluted, so as to vaporise into clusters.
	Inside the spinodal region, fluctuation seeds initiate the process of instability growth.
	The leading disturbance $\lambda^{\prime}$ is the wavelength with the largest growth rate and it induces the development of blobs of size 
$A^{\prime}\approx \rho^{\prime}\,(\lambda^{\prime}/2)^3$.
	If the radial expansion is sufficiently large, those blobs may separate into fragments of corresponding size, situated approximately in the region of Neon and Oxygen.
	On the contrary, when the radial expansion is not large enough, the blobs coalesce into fragments of larger size than $A^{\prime}$.
	Such a scenario is illustrated in fig.~\ref{fig_spinodal} for a bombarding energy which slightly exceeds the threshold between compound nucleus formation and multifragmentation. In this situation we simulate a heavy-ion collision collision leading to a spinodal behaviour: 
at around 100 fm/c the system already presents a mottling pattern; at around 150 fm/c a large number of blobs of comparable size $A^{\prime}$ are the signature of the spinodal process (in the specific event simulated in fig.~\ref{fig_spinodal} about ten blobs of similar size can be counted!); at later times, the spinodal blobs merge together smearing the spinodal signature and forming a smaller number of intermediate-mass fragments (four fragments in this specific event in fig.~\ref{fig_spinodal}), which may even lose their symmetry in size.

\section{Conclusions}

	This work presented the main steps to validate a transport model applied to a fermionic system in presence of mechanical instabilities.
	In particular, the amplitude of the fluctuations are studied as a function of time in correspondence with the properties of the nuclear interaction.
	A transport approach constructed by requiring to satisfy the dispersion relation proves to be suited for the description of nuclear multifragmentation.
	The dynamical approach presented in this work does not require any thermodynamic hypothesis (equilibration for instance), but the characteristic thermodynamic features of multifragmentation, like the occurrence of a nuclear liquid-gas phase transition, are obtained as a result of the transport dynamics~\cite{Napolitani2013}: this finding makes the present dynamical description and alternative statistical approaches for multifragmentation mutually consistent.

%
%
%
%

\end{document}